\documentstyle[12pt,titlepage]{article}

\titlepage

\begin{document}

\vspace{5cm}

\begin{center}
{\large \bf CRITICAL BEHAVIOUR  OF 3D SYSTEMS WITH LONG-RANGE
CORRELATED QUENCHED DEFECTS}
\vspace{1cm}

{V.V. Prudnikov and A.A.Fedorenko\\
Dept. of Theoretical Physics, Omsk State University\\
55a, Pr. Mira, 644077, Omsk, Russia\\
(July 24, 1999)}
\end{center}

\vspace{1cm}

\begin{center}
{\large \bf Abstract}
\end{center}

\vspace{1cm}

A field-theoretic description of the critical behaviour of systems with
quenched defects obeying a power law correlations $\sim |{\bf x }|^{-a}$
for large separations ${\bf x}$ is given. Directly for three-dimensional
systems and different values of correlation parameter $2\leq a \leq 3$
a renormalization analysis of scaling function in the two-loop approximation
is carried out, and the fixed points corresponding to stability
of the various types of critical behaviour are identified. The obtained
results  essentially differ from  results evaluated by double
$\varepsilon, \delta$ - expansion.
The critical exponents in the two-loop approximation are calculated with
the use of the Pade-Borel summation technique.

\vspace{1cm}


\newpage
In last years, much effort has been devoted to investigation of
the critical behaviour of solids containing quenched defects. In most
papers considerations have been restricted to the case of point defects
with small concentrations so that the defects and corresponding random
fields have been assumed to be Gaussian distributed and $\delta$ - correlated.

For the first time in work of Weinrib and Halperin (WH) \cite{1} have been
offered the model of the critical behaviour of disordered system in which the
correlation function of the random local transition temperature
$g({\bf x}-{\bf y})=<<T_c({\bf x})T_c({\bf y})>> - <<T_c({\bf x})>>^2$
falls off with distance as a power law $\sim |{\bf x - y}|^{-a}$.
It was shown that for $a\geq d$ long-range correlations are irrelevant and
the usual short-range Harris criterion \cite{2} $2 - d\nu_o = \alpha_o > 0$
of influence of $\delta$ - correlated point defects is realized, where $d$ is
the spatial dimension, $\nu_o$ and $\alpha_o$ are the correlation-length and
the specific-heat exponents of the pure system. For $a < d$ it was
established the extended criterion $2 - a\nu_o > 0$ of disorder influence
on the critical behaviour. As a result, a wider class of disordered systems,
but not only three-dimensional Ising model with $\delta$-correlated point
defects, can be characterized by new type of critical behaviour. So, for
$a < d$ it was discovered a new long-range (LR) disorder stable fixed
point (FP) of the renormalization group recursion relations for systems with
number of components of the order parameter $m \geq 2$. The critical exponents
were calculated in the one-loop approximation using a double expansion in
$\varepsilon = 4 - d \ll 1$ and $\delta = 4 - a \ll 1$. In the case $m = 1$
the accidental degeneracy of the recursion relations in the one-loop
approximation did not permit to find LR disorder stable FP, but it was
predicted for $\delta > \delta_c = 2(6\varepsilon/53)^{1/2}$ a change
in critical behaviour of the model from short-range (SR)
to LR correlation type. Korzhenevskii, Luzhkov and Schirmacher \cite{3}
have proved the existence of the LR disorder stable FP for the one-component
WH model and found the characteristics of this type of critical behaviour.
Also they have considered very interesting model of the critical behaviour of
crystals with LR correlations caused by point defects with degenerate
internal degrees of freedom \cite{3,4}.

The models with LR correlated quenched defects present both doubtless
theoretical interest from possibility of prediction a new types of the
critical behaviour in disordered systems and experimental interest from
possibility of realization RL correlated defects in the orientational
glasses \cite{5} and disordered solids containing defects
of fractal-like type \cite{3}.
However, numerous investigations of pure and disordered systems performed
with the use of the field-theoretic approach show that the predictions made in
the one-loop approximation, especially on the basis of the $\varepsilon$ -
expansion, can differ strongly from the real critical behaviour [6-9].
Therefore, the map  of regions with the various types of critical behaviour
received for WH model on the basis of $\varepsilon, \delta$ - expansion \cite{1}
(figure 1(a)) may be not corresponding to the critical behaviour of the
three-dimensional WH model for different values of $m$ and $a$. In this case
the results for the models with LR correlated defects received with use of
$\varepsilon, \delta$ - expansion [1,3,4,10-12] must be corrected.
To shed light on this question and to determine more accurately the
dependence of the critical behaviour on the number of components of the
order parameter $m$ and the values of correlation parameter $a$, we have
constructed a field-theoretical description of the three-dimensional WH model
in the two-loop approximation for the values of $a$ in the range
$2\leq a \leq 3$.

The effective Hamiltonian of WH model after using the replica trick is given
by
\begin{eqnarray}
 H_{\rm eff}=\sum_{\alpha=1}^{n}\int d^dx[\frac{1}{2}(r_0\phi_{\alpha}^2+
 (\nabla\phi_{\alpha})^2)+ \frac{u_{0}}{4!}(\phi_{\alpha}^2)^2]-
 \sum_{\alpha,\beta=1}^{n}\int d^dx d^dy g({\bf x}-{\bf y})
 \phi_{\alpha}^2({\bf x})\phi_{\beta}^2({\bf y})
\end{eqnarray}
where $\phi_{\alpha}^2=\sum_{i=1}^{m}\phi_{i \alpha}^2$,\ \
$\phi_{i \alpha}$ is ($n\times{m}$)-component order parameter. The properties
of the original disordered system are obtained in the replica number limit
$n\rightarrow 0$. The Fourier transformation of the interaction vertex
$g(x) \sim x^{-a}$  gives $g(k)=v_0+w_0k^{a-d}$ for small $k$.
$g(k)$ must be positive definite, therefore if $a>d$, then the $w$ term is
irrelevant, $v_0\geq0$ and  $H_{\rm eff}$ (1) corresponds to model with SR -
correlated defects, while if $a<d$, then the $w$ term is dominant at small
$k$ and $w_0\geq 0$.

As is known, in the field-theoretic approach \cite{13} the
asymptotic critical behaviour of systems in the fluctuation region are
determined by the Callan-Symanzik renormalization-group equation for the
vertex parts of the irreducible Green's functions. To calculate the
$\beta$ functions and the critical exponents as functions of the renormalized
interaction vertices $u$, $v$ and $w$ (scaling $\gamma$ functions)
appearing in the renormalization-group equation, we used the standard method
based on the Feynmann diagram technique and the renormalization procedure
\cite{14}. The three types of interactions can be represented
graphically as in figure 2(a). When we considered a diagrammatic representation
of two-point vertex function $\Gamma^{(2)}$, three types of four-point
vertex functions $\Gamma^{(4)}_i$ and two-point with the $\phi^2$ insertion
vertex function $\Gamma^{(1,2)}$ in the two-loop approximation the diagrams
were integrated numerically in $d = 3$ and with values of parameter $a$
determining momentum dependence of the $w$ interaction in the range
$2\leq a \leq 3$ with changes through the step $\Delta a = 0.01$. Unlike the
works using $\varepsilon, \delta$ - expansion we took into consideration
the graphs of the form (figure 2(b)), contributions of which are increased
when the values $a$ are removed from $a=3$.

As a result, we obtained the $\beta$ and $\gamma$ functions in the two-loop
approximation in the form of the expansion series in renormalized vertices
$u$, $v$ and $w$. Because of impossibility in short notes to present the
coefficients of these series for different values of $a$ we give here
the obtained $\beta$ and $\gamma$ functions only for $a=2$ (the case with
$a=2$ corresponds to system of straight lines of impurities or straight
dislocation lines of random orientation in a sample):

\begin{equation}
\begin{array}{rl}
  & \beta_u(u,v,w)=-u+u^2-\frac{3}{2}uv-1.901416uw-\frac{4(41m+190)}{27(m+8)^2}u^3+\frac{2(25m+131)}{27(m+8)}u^2v- \\
  & \ \ \ \ \ \ \ \ \ \ \ \ \ \ \ \ -\frac{185}{216}uv^2+\frac{(1.230378m+6.713002)}{m+8}u^2w-0.312654uw^2-1.193479uvw, \\
  & \beta_v(u,v,w)=v+v^2+\frac{3}{2}w^2+1.901416vw-\frac{2(m+2)}{(m+8)}uv+\frac{95}{216}v^3+0.488229w^3-\\
  & \ \ \ \ \ \ \ \ \ \ \ \ \ \ -\frac{50(m+2)}{27(m+8)}uv^2-1.974883\frac{(m+2)}{(m+8)}uw^2+\frac{92(m+2)}{27(m+8)^2}u^2v+0.806375vw^2+\\
  & \ \ \ \ \ \ \ \ \ \ \ \ \ \ +0.839125v^2w-1.939086\frac{(m+2)}{(m+8)}uvw,   \\
  & \beta_w(u,v,w)=2w+0.628176w^2+\frac{1}{2}vw-\frac{2(m+2)}{(m+8)}uw-0.1528w^3+\frac{92(m+2)}{27(m+8)^2}u^2w+ \\
  & \ \ \ \ \ \ \ \ \ \ \ \ \ \ +\frac{23}{216}v^2w+0.090516\frac{(m+2)}{(m+8)}uw^2-0.022629vw^2-\frac{23(m+2)}{27(m+8)}uvw, \\
  & \gamma_{\phi}(u,v,w)=0.004222w+\frac{8(m+2)}{27(m+8)^2}u^2+\frac{1}{108}v^2+0.056893w^2-\frac{2(m+2)}{27(m+8)}uv- \\
  & \ \ \ \ \ \ \ \ \ \ \ \ \ \ -0.315823\frac{(m+2)}{(m+8)}uw+0.078956vw, \\
  & \gamma_{\phi^2}(u,v,w)=-\frac{m+2}{m+8}u+\frac{1}{4}v+0.31831w+\frac{2(m+2)}{(m+8)^2}u^2+\frac{1}{16}v^2-0.019507w^2- \\
  & \ \ \ \ \ \ \ \ \ \ \ \ \ \ -\frac{(m+2)}{2(m+8)}uv-0.270565\frac{(m+2)}{(m+8)}uw+0.067641vw.
\end{array}
\end{equation}
The series (2) are normalized by a standard change of variables \cite{7,8}
$u \rightarrow 6u/(m+8)J$, $v \rightarrow v/32J$, $w \rightarrow w/32J$, so
that the coefficients of the terms $u, u^2$ and $v, v^2$ in $\beta_u$ and
$\beta_v$ become 1 in modulus, where $J = \int d^{d}q/(q^2+1)^2$ is the
one-loop integral.

The nature of the critical behaviour is determined by the existence of a
stable FP satisfying the system of equations
\begin{equation}
\beta_{i}(u^*,v^*,w^*)=0\ \ \ \ \ \ \ \ \ \ \ \ \ \   (i=1,2,3).
\end{equation}
It is well known, that perturbation series are asymptotically convergent,
and the vertices describing the interaction of the order parameter
fluctuations in the fluctuating region $r\rightarrow 0$ are large enough so
that expressions (2) can't be used directly. For this reason, to extract the
required physical information from the obtained expressions, we employed the
Pade-Borel approximation of summation of asymptotically convergent series
extended to the multiparameter case \cite{9,15}. We used the [2/1] approximant
to calculate the $\beta$ functions in the two-loop approximation.

However, the analysis of the series coefficients for $\beta_w$ function has
shown that the summation of this series is fairly poor, which resulted in
absence of FP with $w^* \neq 0$, for example, in the case $m = 1$ for
$a < 2.93$, in the case $m = 2$ for $a < 2.67$  and so on.
Dorogovtsev has found the symmetry of
the scaling function for WH model in relation to the transformation
$(u, v, w) \rightarrow (u, v, v + w)$ \cite{10} which gives the possibility to
investigate the problem of FP existence with $w^* \neq 0$ in the variables
$(u, v, v + w)$. In this case our investigations have shown the existence
of FP with $w^* \neq 0$ in the whole region where the parameter $a$ changes.

We have found three types of FP's in the physical region of parameter space
$u^*, v^*, v^* + w^* \geq 0$ for different values of $m$ and $a$.
The type I corresponds to FP of pure system $(u^* \neq 0, v^*, w^* = 0)$,
the type II is SR-disorder FP $(u^*, v^* \neq 0, w^* = 0)$ and
the type III corresponds to LR-disorder FP's $(u^*, v^*, w^* \neq 0)$.
The type of the critical behaviour of this disordered system for each value
of $m$ and $a$ is determined by stability of corresponding FP.
The requirement that the fixed point be stable reduces to the condition that
the eigenvalues of the matrix
\begin{equation}
B_{i,j}=\frac{\partial\beta_i(u_1^*,u_2^*,u_3^*)}{\partial{u_j}}
\end{equation}
lie in the right-hand complex half plane.

Values of the stable FP's obtained for the most interesting values of the
number of order-parameter components $m$ and $2\leq a \leq 3$ are presented
in table 1. As one can see from this table, for Ising model $(m = 1)$ the
LR-disorder FP is stable for values of $a$ in the whole investigated range.
The additional calculations for $3 < a < 4$ have shown that only FP II
is stable in this range. For $a = 3$  FP values for vertices $u$ and
$g(k)$ are equal $u^* = 2.383\,38$, $g^* = v^* + w^* = 0.551\,64$
and correspond to SR-disordered Ising model FP, although $w^* \neq 0$.
Like that, for $m = 1$ and $a = 3$ the LR disorder is marginal,
and the critical behaviour of WH model, as of SR-disordered Ising model,
is characterized by the same critical exponents (table 2).
The critical behaviour of the XY-model $(m = 2)$ is determined by
the LR-disorder FP for $a \leq 2.96$ and the SR-disorder FP for
$a > 2.96 $. The Heisenberg model $(m = 3)$ is characterized by the
change of the critical behaviour types from the LR-disorder type (III)
for $a \leq 2.85$ to the pure type (I) for $a > 2.85 $.
Figure 1(b) shows regions of the various types of critical behaviour
of the WH model, which we obtained in the two-loop approximation.
The large change in the picture indicates that the correspondence between
WH results and our calculations in the two-loop approximation is weak.

However, the results, which we received for disordered XY-model, must be
corrected. We think that in the higher field-theory orders of
approximation $k$ the critical behaviour of  XY-model will be determined
by the FP of pure type (I) for $a_{c}^{(k)}< a$, but not by the SR-disorder
FP (II), obtaining in the two-loop order. Here, $a_{c}^{(k)}$ is marginal
value for $a$ in $k$-th order of approximation, for which disorder is
irrelevant ($a_{c}^{(6)} \simeq 2/\nu_o = 2.99$ with $\nu_o = 0.669$ \cite{16}
for $m = 2$). The two facts indicate this, such as the weak stability
of the SR-disorder FP revealed for $2.96 < a < 4 $ and that the
$a_{c}^{(2)} = 3$ for $m_{c} = 2.0114$. In the higher orders of approximation
the marginal value of $m_{c}$ can be found with the use of the Harris
criterion \cite{2} $d\nu_o(m_c) - 2 = 0$, and such as $\nu_o = 0.669$
\cite{16} for $m = 2$, then $m_c < 2$. Therefore, we think that the corrected
picture of the regions of various types of the critical behaviour of
the model with  LR-correlated defects will be represented by figure 1(c).

Finally, we have calculated the static critical exponents for the WH model
(table 2), received from the resummed by the generalized Pade-Borel method
$\gamma$ functions in the corresponding stable FP's
$\eta = \gamma_{\phi}(u^*, v^*, w^*)$,
$\nu=[2+\gamma_{\phi^2}(u^*, v^*, w^*)-\gamma_{\phi}(u^*, v^*, w^*) ]^{-1}$.
Also, we have found by the method of the work \cite{8} the dynamic scaling
function $\gamma_{\lambda}$ and calculated the values of the dynamic
exponent $z = 2 + \gamma_{\lambda}(u^*, v^*, w^*)$ on the basis of
the resummed $\gamma_{\lambda}$ function (table 2). As example, for
$a = 2$	the received $\gamma_{\lambda}$ function is given by
\begin{equation}
\begin{array}{rl}
  & \gamma_{\lambda}(u,v,w)=\frac{1}{4}v+0.314088w+0.226777\frac{(m+2)}{(m+8)^2}u^2+\frac{23}{432}v^2-0.0764w^2- \\
  & \ \ \ \ \ \ \ \ \ \ \ \ \ \ \ -0.092593\frac{(m+2)}{(m+8)}uv+0.123604\frac{(m+2)}{(m+8)}uw-0.011315vw.
\end{array}
\end{equation}

The comparison  of the exponent $\nu$ values and ratio $2/a$ from table 2
shows the violation of supposed in \cite{1} on the basis of some heuristic
arguments as exact the relation $\nu = 2/a$. The revealed difference is
caused by the use in our work of more accurate field-theoretic description
in the higher order of approximation for three-dimensional system directly
together with methods of series summation. Also, these distinctions can be
explained by the application for calculations of the concrete numerical
values of parameter $a$ and the taking into consideration
the graphs of the form (figure 2(b)), thrown away when $\varepsilon, \delta$ -
expansion is used, but contributions of which are increased when
the values $a$ are removed from $a=3$. Of course, there are errors of the
present consideration determined by the accuracy of series summation for
the $\beta$ and $\gamma$ functions. However, the comparison of exponents
values for SR-disorder Ising model, calculated with the use Pade-Borel
method in \cite{6} and \cite{7} in the two-loop and four-loop approximations
accordingly, shows that their differences not more than 0.02. In the same
time, in our work $\nu - 2/a$ depends on the values of $a$ and $m$ and has
the value 0.284, as example, for $a=2$ and $m=1$, that is considerably
larger.

In closing, we hope that the features of the critical behaviour of the WH
model revealed in our paper will stimulate the organization of experimental works
in real disordered systems with long-range correlated defects like
the orientational glasses and solids with defects of fractal-like type.
Also, the computational methods can be applied for simulating disordered
systems with straight lines of impurities of random orientation in a sample
($a=2$). The received values of exponents can be used for explanation of
the results of computer simulation of the three-dimensional disordered Ising
model \cite{17} at impurity concentrations between the threshold of impurity
percolation and the spin-percolation threshold, in which the fractal-like
behaviour of impurity extended structures and the competition between
impurity-percolating and spin-percolating clasters are possible.

\newpage
\begin{center}
{\bf Figure and table captions}
\end{center}

\hspace{-8mm} Figure 1:

\vspace{2mm}

Regions of the various types of critical behaviour, which have been determined
(a) in \cite{1} on the basis of the double $\varepsilon, \delta$ - expansion;
(b) in the present paper with use of the field theoretic description
in two-loop approximation for three-dimensional WH model;
(c) in the present paper with taking into consideration of the higher orders
of approximation.

\vspace{5mm}

\hspace{-8mm} Figure 2:

\vspace{2mm}

Graphs (а) that correspond to vertices $u$, $v$ and $w$;
(b) that are taken into consideration in addition as compared
with works, using $\varepsilon, \delta$ - expansion,\ \
\begin{picture}(15,10)
\multiput(0,0)(0,0.1){100}{\line(1,0){10}}
\end{picture}
corresponds to vertices $u$, $v$ and $w$.

\vspace{5mm}

\hspace{-8mm} Table 1:

\vspace{2mm}

Stable fixed points  of the 3D WH model from two-loop expansions.

\vspace{5mm}

\hspace{-8mm} Table 2:

\vspace{2mm}

Critical exponents of the 3D WH model from two-loop expansions.

\newpage
\begin{table}
\hspace{120mm} {Table 1}
\vspace{5mm}
\begin{center}
\begin{tabular}{|c|c|c|c|c|c|c|c|} \hline
      &        $n=1$                &           $n=2$            &          $n=3$             \\  \hline
  a   &  \ \ \ \ \ $u^{*}$ \ \ \ \ \ \ $v^{*}$ \ \ \ \ $w^{*}+v^{*}$ & \ \ \ \ \ $u^{*}$ \ \ \ \ \ \ $v^{*}$ \ \ \ \ $w^{*}+v^{*}$ & \ \ \ \ \ $u^{*}$ \ \ \ \ \ \ $v^{*}$ \ \ \ \ $w^{*}+v^{*}$ \\ \hline
 ...  &  ...  \ \ \ \ \ \ \  ... \ \ \ \ \ \ \ ... &  ...  \ \ \ \ \ \ \  ... \ \ \ \ \ \ \ ... &  ...  \ \ \ \ \ \ \  ... \ \ \ \ \ \ \ ... \\
 3.1  & 2.38338   0.55164   0.55164   & 1.56469   0.00416   0.00416  & 1.52097   0.00000   0.00000  \\
 3.0  & 2.38338   0.22293   0.55164   & 1.56469   0.00416   0.00416  & 1.52097   0.00000   0.00000  \\
 2.9  & 2.59804   0.31890   0.68114   & 2.09001   0.11386   0.40038  & 1.52097   0.00000   0.00000  \\
 2.8  & 2.77927   0.40153   0.78299   & 2.17677   0.13536   0.44359  & 1.95770   0.08298   0.34550  \\
 2.7  & 2.94031   0.47487   0.86757   & 2.26778   0.15923   0.48612  & 2.01746   0.09346   0.37004  \\
 2.6  & 3.08645   0.54084   0.93916   & 2.36058   0.18457   0.52633  & 2.08699   0.10922   0.40005  \\
 2.5  & 3.21983   0.60035   0.99972   & 2.49643   0.23442   0.59651  & 2.15585   0.12535   0.42628  \\
 2.4  & 3.34078   0.65374   1.04998   & 2.61818   0.28094   0.65334  & 2.22047   0.14074   0.44651  \\
 2.3  & 3.44813   0.70082   1.08980   & 2.72520   0.32344   0.69760  & 2.30801   0.16910   0.48302  \\
 2.2  & 3.53899   0.74092   1.11825   & 2.81501   0.36115   0.72909  & 2.39298   0.20079   0.51696  \\
 2.1  & 3.60814   0.77263   1.13340   & 2.88305   0.39293   0.74672  & 2.45869   0.22877   0.53759  \\
 2.0  & 3.64687   0.79347   1.13189   & 2.92206   0.41710   0.74843  & 2.49945   0.25161   0.54364  \\  \hline
\end{tabular} \end{center} \end{table}

\newpage
\begin{table}
\hspace{120mm} {Table 2}
\vspace{5mm}
\begin{center}
\begin{tabular}{|c|c|c|c|c|} \hline
     &      &        $n=1$                &           $n=2$            &          $n=3$             \\  \hline
 $a$ &$2/a$ &$\eta$ \ \ \ \ \ \ $\nu$ \ \ \ \ \ $z$&$\eta$ \ \ \ \ \ \ $\nu$ \ \ \ \ \ $z$&$\eta$ \ \ \ \ \ \ $\nu$ \ \ \ \ \ $z$ \\  \hline
 3.1 &      &  0.0327  0.6715   2.1712  &  0.0288  0.6642  2.0000  &  0.0283  0.6960 2.0217  \\
 3.0 &0.6667&  0.0327  0.6715   2.1712  &  0.0288  0.6642  2.0000  &  0.0283  0.6960 2.0217  \\
 2.9 &0.6897&  0.0304  0.6813   2.2120  &  0.0248  0.7141  2.1315  &  0.0283  0.6960 2.0217  \\
 2.8 &0.7143&  0.0270  0.6889   2.2486  &  0.0212  0.7190  2.1510  &  0.0179  0.7600 2.1128  \\
 2.7 &0.7407&  0.0227  0.6950   2.2837  &  0.0166  0.7240  2.1736  &  0.0137  0.7632 2.1269  \\
 2.6 &0.7692&  0.0176  0.7002   2.3184  &  0.0112  0.7692  2.1988  &  0.0084  0.7682 2.1443  \\
 2.5 &0.8000&  0.0118  0.7046   2.3532  &  0.0035  0.7378  2.2338  &  0.0025  0.7727 2.1633  \\
 2.4 &0.8333&  0.0055  0.7083   2.3879  & -0.0050  0.7452  2.2684  & -0.0040  0.7763 2.1827  \\
 2.3 &0.8696& -0.0012  0.7114   2.4215  & -0.0138  0.7513  2.3013  & -0.0125  0.7835 2.2078  \\
 2.2 &0.9091& -0.0081  0.7137   2.4524  & -0.0226  0.7558  2.3301  & -0.0218  0.7905 2.2315  \\
 2.1 &0.9524& -0.0147  0.7151   2.4780  & -0.0307  0.7588  2.3522  & -0.0303  0.7952 2.2514  \\
 2.0 &1.0000& -0.0205  0.7155   2.4949  & -0.0371  0.7599  2.3649  & -0.0370  0.7975 2.2644  \\  \hline
\end{tabular} \end{center} \end{table}

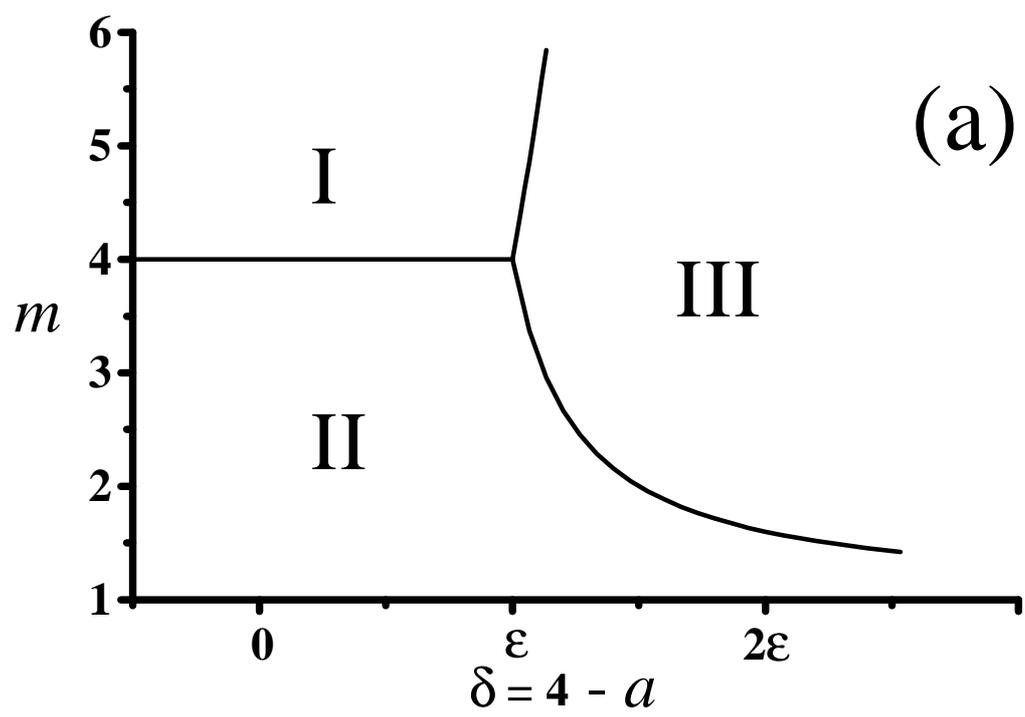
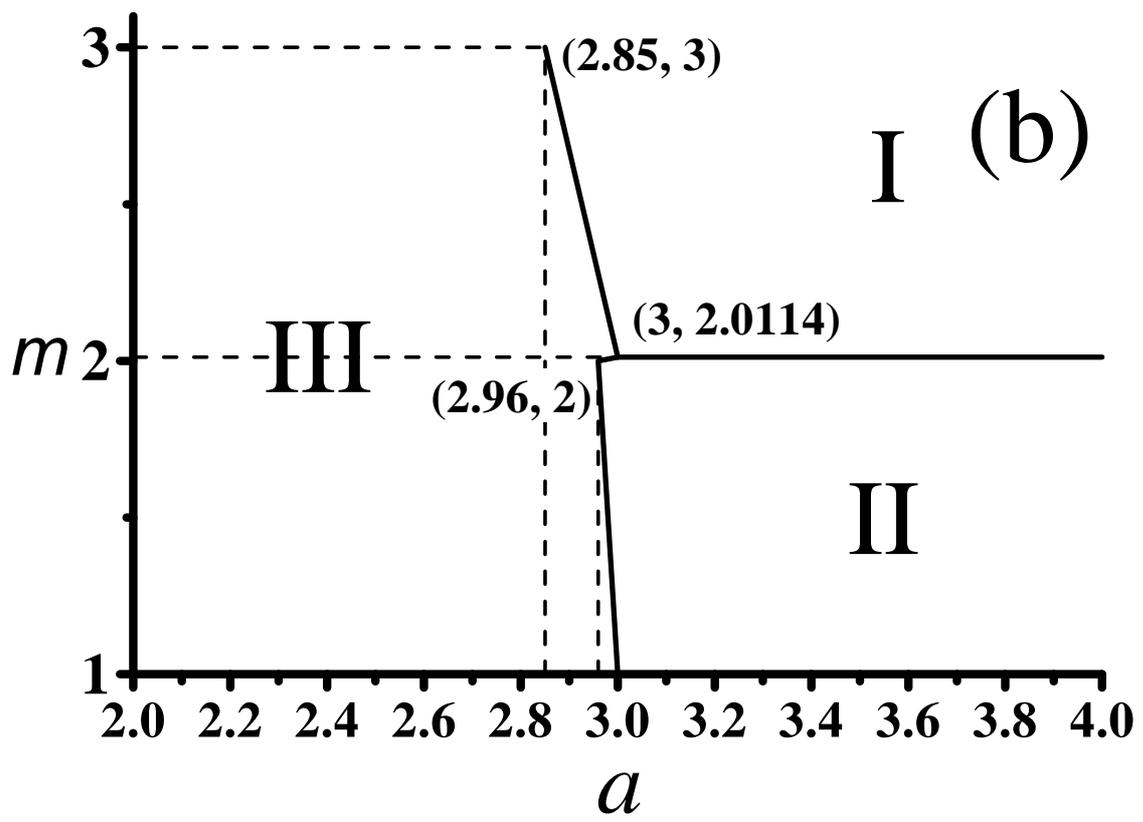
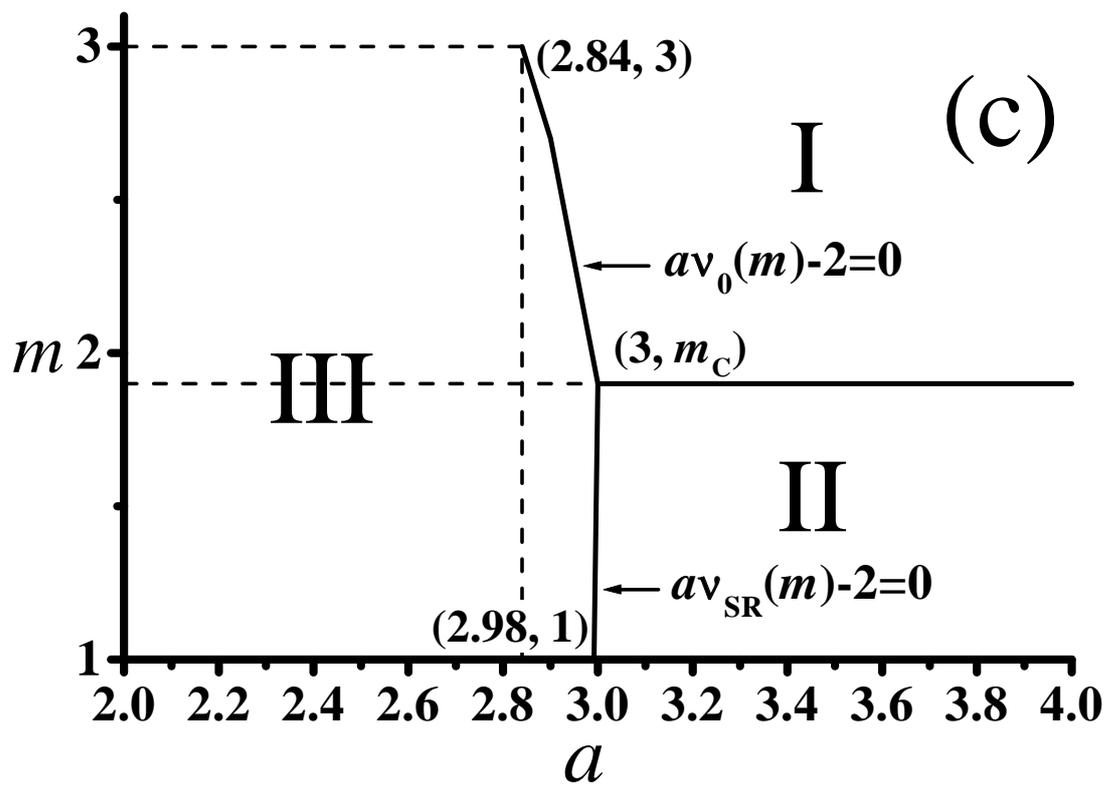
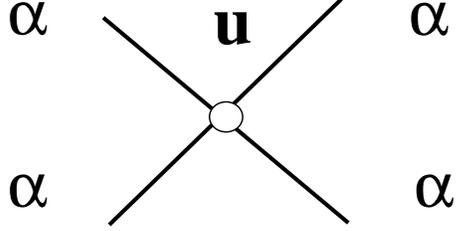
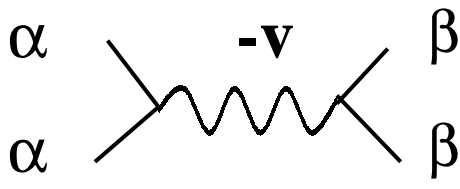
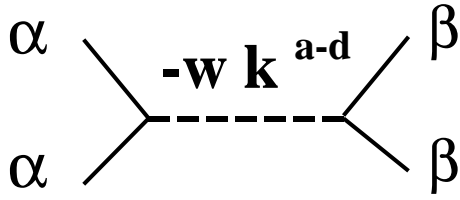
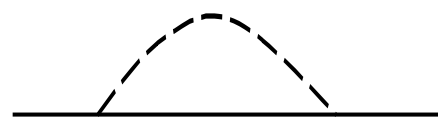
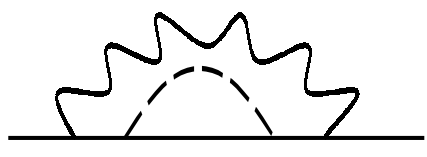
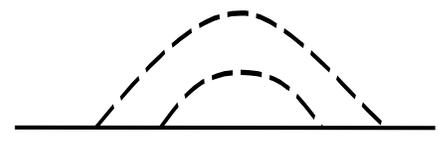
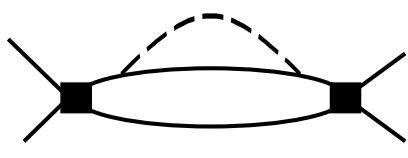
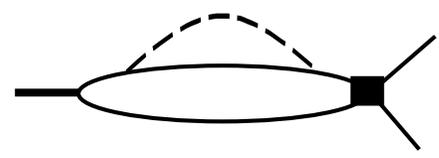
\end{document}